\documentclass{article}
\usepackage{hiph-art}
\usepackage{epsfig}
\usepackage{graphicx}
\usepackage{psfig}
\usepackage{amssymb,amsbsy}
\usepackage{latexsym,epsf,axodraw,pstricks}
\usepackage[english]{babel}
\usepackage{amsmath}
\volnumber{00} \issuenumber{1} \edyear{2005}                             
\frompage{000} \topage{000}                                              
\recrevdate{October 2005}                                              
\def\rightmark{Kinetic and Chemical Equilibration in Scalar $\varphi^4$ Theory}
\def\leftmark{A. Arrizabalaga}
 
\title{{\bf Kinetic and Chemical Equilibration \\
in Scalar {\boldmath$\varphi^4$} Theory}} 
\authors{ 
{Alejandro Arrizabalaga\footnote{Talk based on work in collaboration with J. Smit and A. Tranberg} %
\index{Arrizabalaga, A} 
}\\[2.812mm]
{\normalsize
Theory Department, NIKHEF \\ 
Kruislaan 409, 1098SJ, Amsterdam, The Netherlands\\[0.2ex] 
}}
 
\abstract{
Approximations based on the 2PI effective action are used to investigate the process of equilibration in $\varphi^4$
theory in 3+1 dimensions, both in the symmetric and broken phase. A special emphasis is put on the study of the kinetic and chemical equilibration.
}


\newcommand{\be}{\begin{equation}}
\newcommand{\ee}{\end{equation}}

\newcommand{\bc}{\begin{center}}
\newcommand{\ec}{\end{center}}
\newcommand{\bea}{\begin{eqnarray}}
\newcommand{\eea}{\end{eqnarray}}
\newcommand{\half}{\frac{1}{2}}

\newcommand{\ts}[1]{{\mbox{\scriptsize #1}}}
\newcommand{\tinys}[1]{{\mbox{\tiny #1}}}
\newcommand{\bfig}{\begin{figure}[h]}
\newcommand{\efig}{\end{figure}}

\newcommand{\bi}{\begin{itemize}}
\newcommand{\ei}{\end{itemize}}
\newcommand{\ba}{\begin{align}}
\newcommand{\ea}{\end{align}}
\newcommand{\eref}[1]{(\ref{#1})}

\newcommand{\tpic}[1]{\;\parbox[c]{20pt}{\begin{picture}(20,30)(0,0)
\SetWidth{1.0}\SetScale{1.0} #1 \end{picture}}\;}
\newcommand{\pic}[1]{\;\parbox[c]{30pt}{\begin{picture}(30,30)(0,0)
\SetWidth{1.0}\SetScale{1.0} #1 \end{picture}}\;}

\newcommand{\picb}[1]{\;\parbox[c]{45pt}{\begin{picture}(45,30)(0,0)
\SetWidth{1.0}\SetScale{1.0} #1 \end{picture}}\;}

\makeindex
\begin{document}

\maketitle
\section{Introduction}
To establish the formation of a thermalized quark-gluon plasma during a heavy-ion
collision it is crucial to determine the time scales involved in the equilibration of the initial fireball. The
hydrodynamic description of the experiments suggests early thermalization \cite{Heinz:2001xi}, which seems to contradict traditional
perturbative estimates \cite{Molnar:2001ux}. The resolution of this puzzle has spurred the development of techniques to understand the microscopic dynamical
processes leading to equilibration. 
\par
A powerful formalism to study out-of-equilibrium dynamics from first principles is the 2PI effective action. It provides an exact representation of a given theory in terms of a functional depending solely on the connected one-
and two-point functions. Approximate equations describing the time evolution of these correlators can be derived from a
variational principle on the functional. The main interest of the 2PI effective action as a framework to obtain evolution
equations is the fact that the global symmetries of the theory are respected \cite{Baym:1962sx}. In particular, the evolution equations
guarantee energy conservation, which is an important feature when studying out-of-equilibrium processes.
They also seem to be free of the secular instabilities that plague traditional perturbative approaches.
\par
In recent years, approximations based on the 2PI effective action have been applied succesfully to the study of
far-from-equilibrium dynamics and equilibration. Most studies have been carried out in scalar theories, both in 1+1
\cite{Berges:2000ur,Aarts:2002dj,Cooper:2002qd} and
2+1 dimensions \cite{Juchem:2003bi}, and in Yukawa theory in 3+1 dimensions \cite{Berges:2002wr}. 
The application of 2PI effective action methods
to gauge theories, however, is not straightforward due to a residual dependence on the choice of gauge condition \cite{Arrizabalaga:2002hn}
and the difficulty of implementing properly a numerical
discretisation.  
In this work we extend to 3+1 dimensions previous studies of equilibration in scalar theory, investigating, in addition, the
behavior in the broken phase. More information can be found in \cite{Arrizabalaga:2005tf}. 

\section{2PI Effective action and evolution equations}
For the scalar theory given by the action in the form
\be
S[\varphi]=\int d^4x\,\int d^4y \half\varphi(x)
G_0^{-1}(x,y)\varphi(y)-\frac{\lambda}{4!}\varphi(x)^4
\ee
with $G_0^{-1}(x,y)=(-\partial_x^2-m^2)\delta(x,y)$, the corresponding 2PI effective action is given by \cite{Cornwall:1974vz}
\be
\Gamma_\ts{2PI}[\phi,G]=S[\phi]-\frac{i}{2}\mbox{Tr}\ln
G+\frac{i}{2}\mbox{Tr}\Big[(G_0^{-1}-G^{-1})\cdot G\Big]+\Phi[\phi,G]
\ee
Here $\phi$ and $G$ correspond, respectively, to the full connected one- and two-point correlation functions.
The functional $\Phi$ consists of an inifinite series of closed skeleton diagrams
\be
i\Phi[\phi,G]=
\frac{1}{4}
\tpic{
\Oval(10,22.5)(7.5,7.5)(0)
\Vertex(10,15){2}
\Line(10,15)(0,5)
\Line(10,15)(20,5)
\Line(0,9)(4,5)
\Line(20,9)(16,5)
}
+
\frac{1}{8}
\tpic{
\Oval(10,7.5)(7.5,7.5)(0)
\Oval(10,22.5)(7.5,7.5)(0)
\Vertex(10,15){2}
}
+\frac{1}{12}
\picb{
\GCirc(22.5,15){11.5}{1}
\Line(2,15)(43,15)
\Vertex(11,15){2}
\Vertex(34,15){2}
\Line(0,17)(4,13)
\Line(0,13)(4,17)
\Line(45,17)(41,13)
\Line(45,13)(41,17)
}
+\frac{1}{48}
\pic{
\GCirc(15,15){15}{1}
\Oval(15,15)(6.5,15)(0)
\Vertex(0,15){2}
\Vertex(30,15){2}
}
+\ldots
\ee
where the lines correspond to $G$, the crosses to $\phi$ and the vertices to $-i\lambda$.
\par 
Evolution equations for the one- and two-point functions are obtained by a variational principle on the effective
action $\Gamma_\ts{2PI}$, conveniently formulated using a real-time contour for the time integrations.
One obtains
\begin{align}
	\frac{\delta \Gamma_\ts{2PI}[\phi,G]}{\delta G}&=0\rightarrow
	\delta_\mathcal{C}(x,y)=\int_{\mathcal{C}}d^4z\,G_0^{-1}(x,z)G(z,y)+i\int_{\mathcal{C}}d^4z\,\Sigma(x,z)G(z,y) \\
	\frac{\delta \Gamma_\ts{2PI}[\phi,G]}{\delta \phi}&=0\rightarrow
	\left[\partial^2+m^2\right]\phi(x)=-\frac{\lambda}{6}\phi(x)^3+\frac{\delta
\Phi[\phi,G]}{\delta \phi(x)},
\end{align}
with $\mathcal{C}$ being the integration time contour (see for instance \cite{Landsman:1986uw}). The self-energy
$\Sigma(x,y)$ is given by
\be
\Sigma(x,y)=-2\frac{\delta \Phi[\phi,G]}{\delta G(y,x)}.
\ee 
The exact evolution of $\phi$ and $G$ would be determined by solving these equations including the infinite series of
diagrams in $\Phi[\phi,G]$. This is not possible to do, so one considers approximate evolution equations that result
from taking only a finite set of diagrams, i.e. by truncating $\Phi[\phi,G]$ up to some order in a convenient expansion
parameter. In our study we consider various truncations of the loop expansion of $\Gamma_\ts{2PI}$, summarized in Table
\ref{table:truncations}.
\begin{table}
\begin{tabular}{cc}\hline
	\emph{Truncation} & $i\Phi[\phi,G]$ \\ \hline
Hartree approximation &
$
\parbox[c]{1pt}{\begin{picture}(1,25)(0,0)
\SetWidth{1.0}\SetScale{1.0} 
\end{picture}}
\frac{1}{4}\parbox[c]{20pt}{\begin{picture}(20,20)(0,0)
\SetWidth{1.0}\SetScale{1.0} 
\Oval(10,15)(5,4.5)(0)
\Vertex(10,10){1.5}
\Line(10,10)(2.5,2.5)
\Line(10,10)(17.5,2.5)
\Line(5,2.5)(2.5,5)
\Line(15,2.5)(17.5,5)
\end{picture}}
+\frac{1}{8}
\parbox[c]{20pt}{\begin{picture}(20,20)(0,0)
\SetWidth{1.0}\SetScale{1.0} 
\Oval(10,5)(5,4.5)(0)
\Oval(10,15)(5,4.5)(0)
\Vertex(10,10){1.5}
\end{picture}}
$ \\
Two-loop approximation & 
$
\parbox[c]{1pt}{\begin{picture}(1,25)(0,0)
\SetWidth{1.0}\SetScale{1.0} 
\end{picture}}
\frac{1}{4}\parbox[c]{20pt}{\begin{picture}(20,20)(0,0)
\SetWidth{1.0}\SetScale{1.0} 
\Oval(10,15)(5,4.5)(0)
\Vertex(10,10){1.5}
\Line(10,10)(2.5,2.5)
\Line(10,10)(17.5,2.5)
\Line(5,2.5)(2.5,5)
\Line(15,2.5)(17.5,5)
\end{picture}}
+
\frac{1}{8}
\parbox[c]{20pt}{\begin{picture}(20,20)(0,0)
\SetWidth{1.0}\SetScale{1.0} 
\Oval(10,5)(5,4.5)(0)
\Oval(10,15)(5,4.5)(0)
\Vertex(10,10){1.5}
\end{picture}}
+
\frac{1}{12}\;
\parbox[c]{30pt}{\begin{picture}(30,20)(0,0)
\SetWidth{1.0}\SetScale{1.0}
\GCirc(15,10){7}{1}
\Line(2,10)(28,10)
\Vertex(8,10){1.5}
\Vertex(22,10){1.5}
\Line(0.5,11.5)(3.5,8.5)
\Line(0.5,8.5)(3.5,11.5)
\Line(29.5,11.5)(26.5,8.5)
\Line(29.5,8.5)(26.5,11.5)
\end{picture}}
$ \\
``Basketball'' approximation  & 
$
\parbox[c]{1pt}{\begin{picture}(1,25)(0,0)
\SetWidth{1.0}\SetScale{1.0} 
\end{picture}}
\frac{1}{4}\parbox[c]{20pt}{\begin{picture}(20,20)(0,0)
\SetWidth{1.0}\SetScale{1.0} 
\Oval(10,15)(5,4.5)(0)
\Vertex(10,10){1.5}
\Line(10,10)(2.5,2.5)
\Line(10,10)(17.5,2.5)
\Line(5,2.5)(2.5,5)
\Line(15,2.5)(17.5,5)
\end{picture}}
+
\frac{1}{8}
\parbox[c]{20pt}{\begin{picture}(20,20)(0,0)
\SetWidth{1.0}\SetScale{1.0} 
\Oval(10,5)(5,4.5)(0)
\Oval(10,15)(5,4.5)(0)
\Vertex(10,10){1.5}
\end{picture}}
+
\frac{1}{12}\;
\parbox[c]{30pt}{\begin{picture}(30,20)(0,0)
\SetWidth{1.0}\SetScale{1.0}
\GCirc(15,10){7}{1}
\Line(2,10)(28,10)
\Vertex(8,10){1.5}
\Vertex(22,10){1.5}
\Line(0.5,11.5)(3.5,8.5)
\Line(0.5,8.5)(3.5,11.5)
\Line(29.5,11.5)(26.5,8.5)
\Line(29.5,8.5)(26.5,11.5)
\end{picture}}
+\frac{1}{48}\;
\parbox[c]{30pt}{\begin{picture}(20,20)(0,0)
\SetWidth{1.0}\SetScale{1.0}
\GCirc(10,10){10}{1}
\Oval(10,10)(4,9.5)(0)
\Vertex(0,10){1.5}
\Vertex(20,10){1.5}
\end{picture}}
$ \\ \hline
\end{tabular}
\caption{Truncations of the 2PI effective action.}
\label{table:truncations}
\end{table}
\par
In a real-scalar theory, the two-point functions defined on the time contour $\mathcal{C}$ can be written in terms of
two independent components, which we take as
\be
G_\mathcal{C}(x,y)=F(x,y)-\frac{i}{2}\mbox{sign}_\mathcal{C}(x_0-y_0)\rho(x,y).
\ee
The functions $F$ and $\rho$ contain, respectively, statistical and spectral information about the system. They are real
and satisfy
the symmetry properties $F(x,y)=F(y,x)$ and $\rho(x,y)=-\rho(y,x)$, which make them very convenient for numerical
implementation.
In terms of these functions, the evolution equations look like
\begin{align}
\left[ \partial_x^2+M^2(x)\right]F(x,y)&=\int_{0}^{x_0}dz_0 \int d^3
z\;\Sigma^{\rho}(x,z)F(z,y)-\int_{0}^{y_0}dz_0\int d^3z\;\Sigma^{F}(x,z)\rho(y,z),\nonumber\\
\left[\partial_x^2+M^2(x)\right]\rho(x,y)&=\int_{y_0}^{x_0}dz_0\int
d^3z\;\Sigma^{\rho}(x,z)\rho(z,y),
\label{evolutionequations}\\
\left[ \partial_x^2+M^2(x)\right]\phi(x)&=\frac{\lambda}{3}\phi(x)^3+\int_0^{x_0}
dz_0\int d^3z\;
\widetilde{\Sigma}^\rho(x,z)\phi(z),\nonumber
\end{align}
where $M^2(x)=m^2+\frac{\lambda}{2}\phi(x)^2+\frac{\lambda}{2}F(x,x)$. For the \emph{Basketball approximation}, we have
\begin{align}
\Sigma^F(x,y)&=\frac{\lambda^2}{2}\phi(x)\phi(y) \left[
F^2(x,y)-\frac{\rho^2(x,y)}{4}\right]+\frac{\lambda^2}{6}F(x,y) \left[
F^2(x,y)-\frac{3\rho^2(x,y)}{4}\right],\nonumber\\
\Sigma^\rho(x,y)&=\lambda^2 \phi(x)\phi(y)\big[F(x,y)\rho(x,y)\big]
+\frac{\lambda^2}{6}\rho(x,y)\left[3F^2(x,y)-\frac{\rho^2(x,y)}{4}\right],\label{selfenergies}\\
\widetilde{\Sigma}^\rho(x,z)&=-\frac{\lambda^2}{6}\rho(x,z)\left[3F(x,z)^2-\frac{\rho(x,z)^2}{4}\right].
\nonumber
\end{align}
We see from \eref{evolutionequations} that the evolution of $F$, $\rho$ and $\phi$ is determined by their values at previous
times,
which enter the equation in the form of memory integrals. For the \emph{two-loop approximation}, the memory kernels are given by the
first term in the RHS of \eref{selfenergies}, while for the \emph{Hartree approximation} they vanish alltogether.
\par
To solve set of coupled evolution equations \eref{evolutionequations} we need to specify a set of initial
conditions. We consider a
spatially homogeneous situation so that the correlators can be written in terms of their mode functions
$F_\mathbf{k}(t,t^\prime)$ and $\rho_\mathbf{k}(t,t^\prime)$. The initial conditions are then given by the values and
derivatives of the mode functions $F$, $\rho$ and $\phi$ at initial time. In the case of the spectral function $\rho$,
these are specified by the property
\be
\rho_{\mathbf{k}}(t,t)=0,\quad \partial_,
\rho_{\mathbf{k}}(t,t^\prime)\big|_{t=t^\prime}=1.
\ee
The initial conditions for $F$ are taken of the form
\begin{align}
F_{\mathbf{k}}(t,t^\prime)\big|_{t=t^\prime=0}&=\frac{1}{\omega_{\mathbf{k}}}\left[n_{\mathbf{k}}+\half\right],\nonumber\\ 
\partial_t F_{\mathbf{k}}(t,t^\prime)\big|_{t=t^\prime=0}&=0, \label{icF}\\
\partial_t \partial_{t^\prime}
F_{\mathbf{k}}(t,t^\prime)\big|_{t=t^\prime=0}&=\omega_{\mathbf{k}}\left[
n_{\mathbf{k}}+\half \right] \nonumber,
\end{align}
which are identical to the equilibrium free theory case, but with some arbitrary distribution function $n_\mathbf{k}$.
\par
The evolution equations \eref{evolutionequations} are solved numerically by discretizing the theory on a space-time lattice with
spacings $a_t\ll a$. The lattice provides a cut-off and regularizes the ultraviolet divergences present in the continuum
limit, which are to be dealt with by renormalization. The continuum renormalization of approximations based on the 2PI
effective action has been recently studied in detail (see for instance \cite{Berges:2005hc}). For our purpose it is sufficient to use
an approximate renormalization that ensures that the relevant length scales in the simulations are larger than the
lattice spacing $a$. For all the truncations of table \eref{table:truncations} we take a perturbative version of the
renormalization needed just for the Hartree case.

\section{Symmetric Phase}
\begin{figure}
	\begin{minipage}{.5\textwidth}
		\includegraphics[clip=true,width=.9\textwidth]{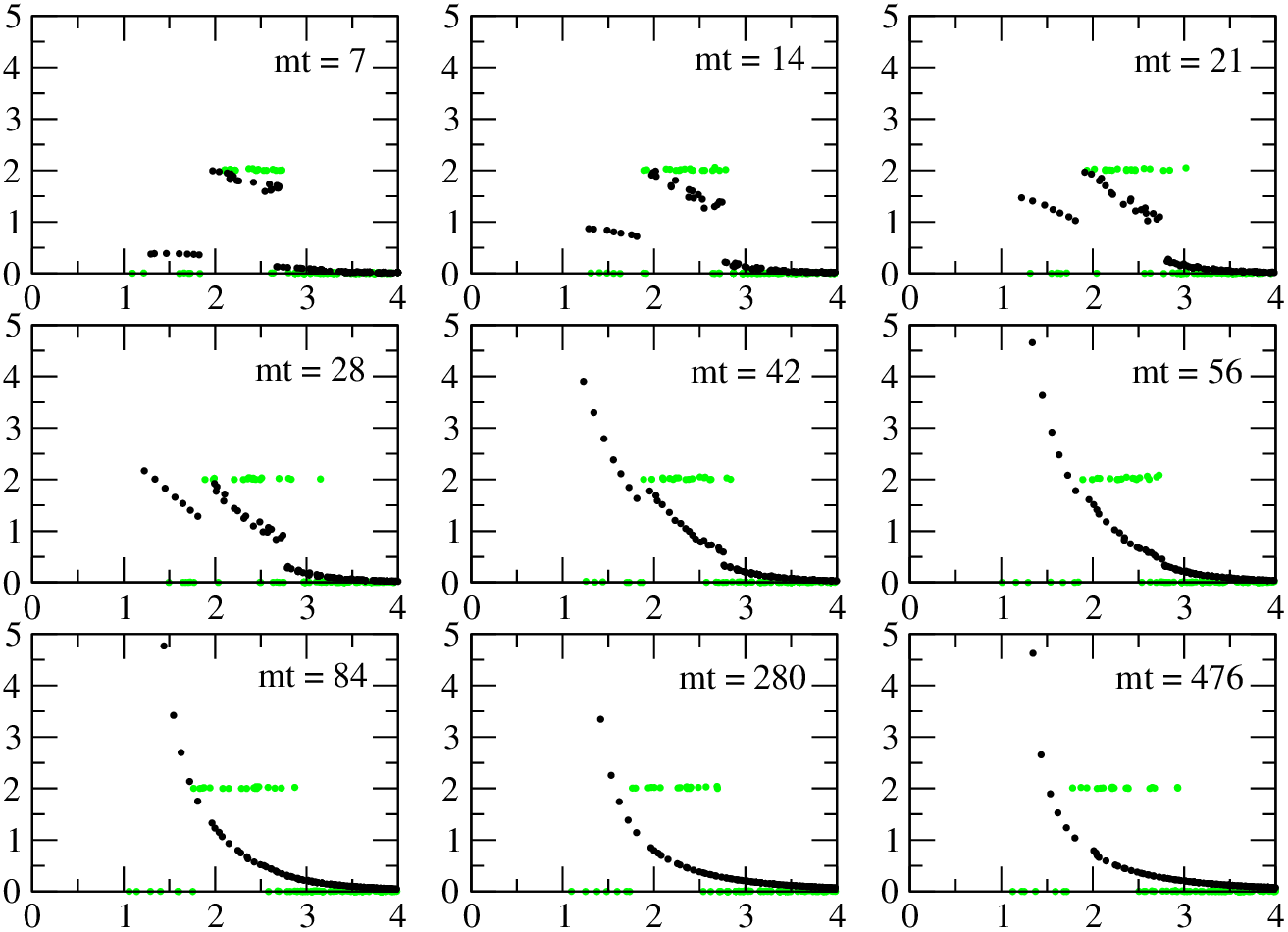}
	\end{minipage}
	\begin{minipage}{.5\textwidth}
		\includegraphics[clip=true,width=.9\textwidth]{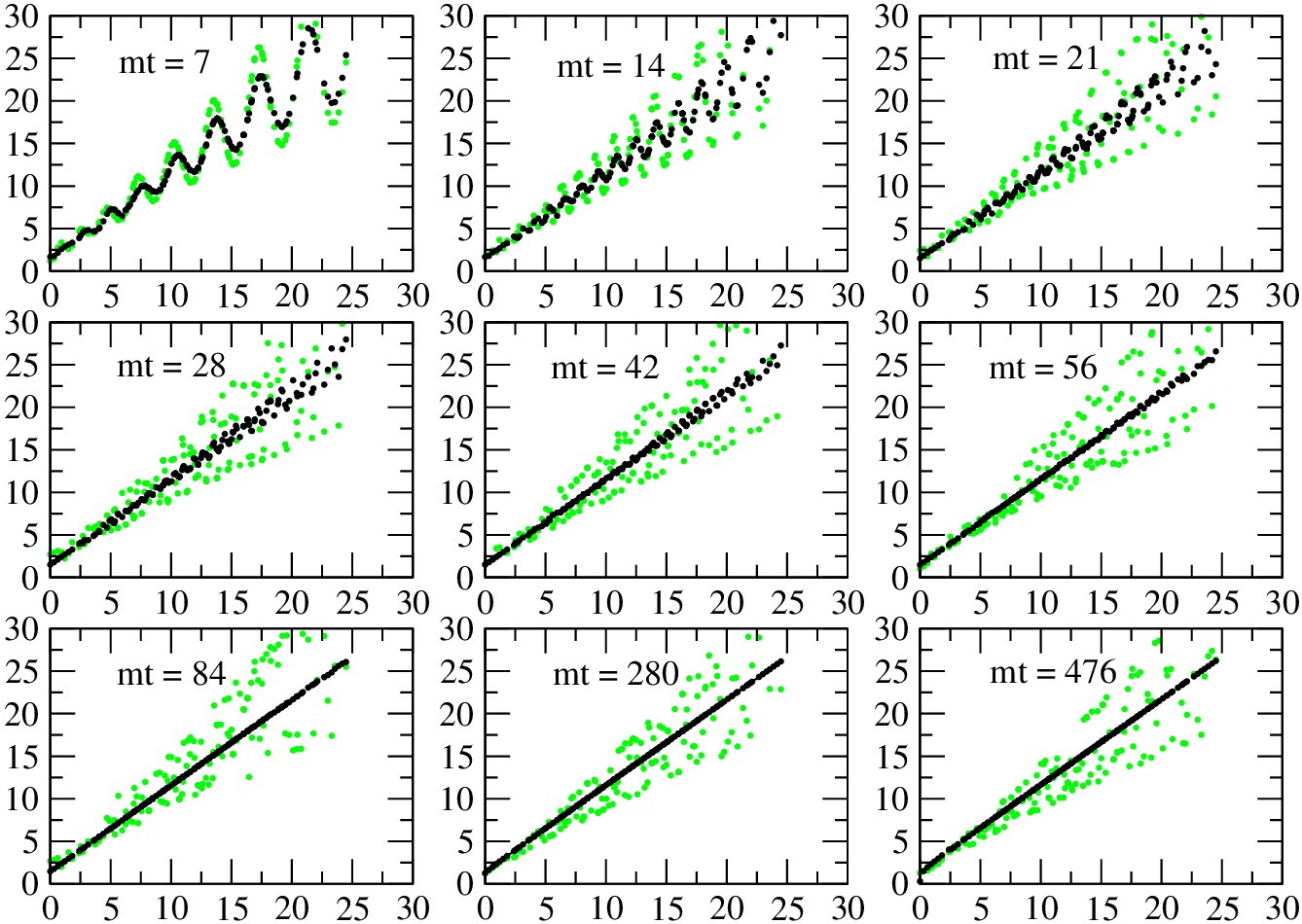}
	\end{minipage}
\caption{Evolution in time of the occupation numbers, $n_{\mathbf{k}}$
\textit{vs.~}$\omega_{\mathbf{k}}$ (left), and dispersion relation $\omega_{\mathbf{k}}^2$
vs.~${\mathbf{k}}^2$ (right), for an initial condition with
$(\eta,\mathbf{k}^2_\tinys{min},\mathbf{k}^2_\tinys{max})=(2,2.04m^2,6.12m^2)$ (T1). We display the results of the ``Basketball''
(black 
dots) and the Hartree approximation
(green/grey 
dots).}
\label{figure1}
\end{figure}

\begin{figure}
\begin{minipage}[t]{.49\textwidth}
	\centerline{\includegraphics[clip=true,width=.9\textwidth]{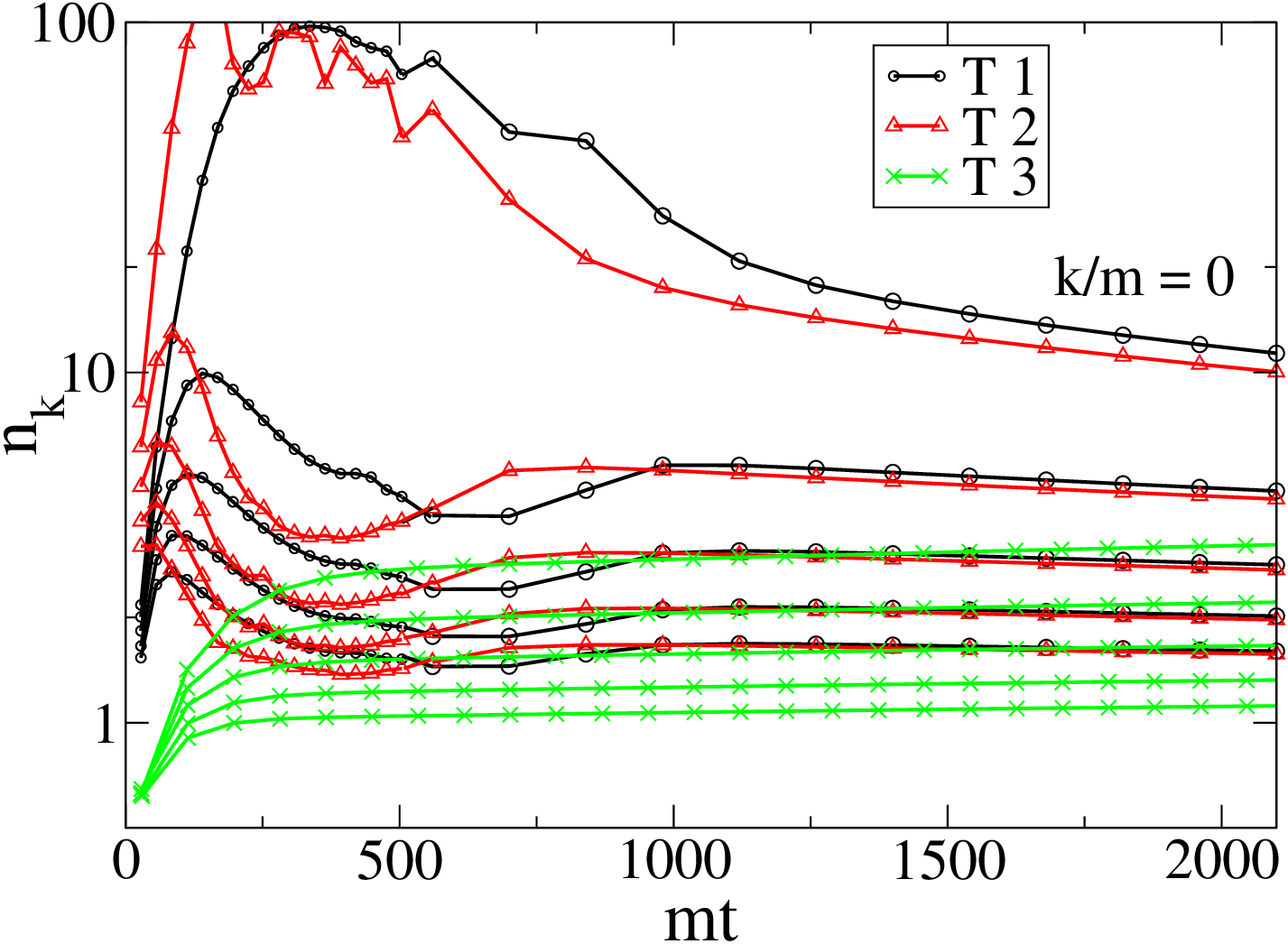}}
\caption{Evolution of individual modes for various initial conditions with same energy density (T1,T2 and T3).}
\label{figure2}
\end{minipage}
\begin{minipage}[t]{.49\textwidth}
	\centerline{\includegraphics[clip=true,width=.9\textwidth]{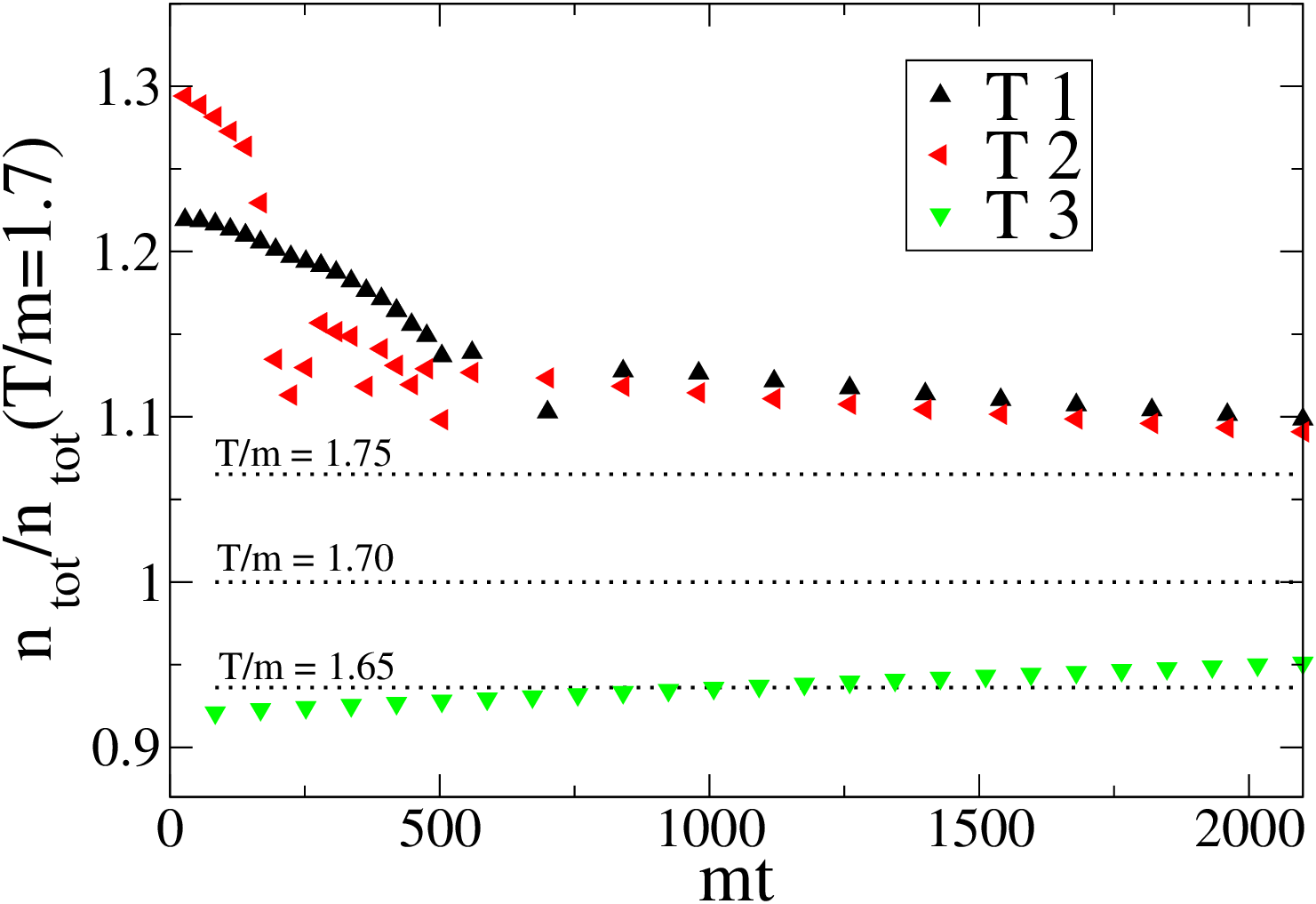}}
\caption{Evolution of total particle number density $n_\tinys{tot}$ for (T1,T2 and T3)}
\label{figure3}
\end{minipage}
\end{figure}

We consider first the evolution of the system in the symmetric phase. We initialize the system far from
equilibrium with a distribution of the form
\be
n_{\mathbf{k}}=\eta\,\Theta({\mathbf{k}}^2_\ts{max}-{\mathbf{k}}^2)\Theta({\mathbf{k}}^2-{\mathbf{k}}^2_\ts{min}),
\label{distro}
\ee
which means that only modes with momenta in the range ${\mathbf{k}}_\ts{min}^2<{\mathbf{k}}^2<{\mathbf{k}}^2_\ts{max}$
are occupied. In the following we shall consider three diferent distributions: T1, T2 and T3, all with identical energy
and T1 and T2 with similar initial particle number. The numerical evaluation of the evolution equations \eref{evolutionequations} is performed on a $16^3$
spatial lattice with spacing $am=0.7$ (in units of the renormalized mass $m$), time spacing $a_t m=0.07$ and coupling
$\lambda=6$.\footnote{The memory integrals in \eref{evolutionequations} are cut to a maximum time extent
$mt_\tinys{cut}\approx 28$. Contributions to the kernels from previous times were checked to be negligible.} 
To monitor the evolution of the system towards equilibrium we study the behavior of an effective quasiparticle distribution function
and a dispersion relation, defined as
\begin{align}
n_{\mathbf{k}}(t)+\half&=c_\mathbf{k}\sqrt{\partial_t \partial_{t^\prime}
F_{\mathbf{k}}(t,t^\prime)\big|_{t=t^\prime}\,F_{\mathbf{k}}(t,t)},\label{particlenumber}\\
\omega_{\mathbf{p}}(t)&=\sqrt{\partial_t \partial_{t^\prime}
F_{\mathbf{k}}(t,t^\prime)\big|_{t=t^\prime}\,/\,F_{\mathbf{k}}(t,t)}.
\label{particleenergy}
\end{align}
The evolution of the distribution $n_{\mathbf{k}}(t)$ and frequency $\omega_{\mathbf{p}}(t)$ is shown in Fig.~\ref{figure1} for the
Hartree and Basketball approximations. In the Hartree case there is no inter-mode communication (no scattering), so the system does not
equilibrate. In the Basketball case, the distribution and dispersion relation approach equilibrium profiles around
$mt\approx 80$. 
To find out if the system has indeed reached a universal equilibrium state, we study the evolution of
individual modes for different initial conditions with identical energy density (see Fig.~\ref{figure2}). One can see
that the individual modes become roughly equilibrated at $mt\approx 1000$. Further on there is practically no inter-mode
communication, so it is reasonable to say that \emph{kinetic equilibration} has been established.
There is still, however, a slow drift towards the complete equilibrium state, for which the mode distributions should agree for
the various initial conditions. As one can see in Fig.~\ref{figure2}, the evolution from the initial conditions $T1$
and $T3$ differs substantially even at $mt\sim 2000$, so the system is still far from complete equilibrium. The reason
for the slow drift towards full equilibrium at later stages seems to be the fact that the system is not chemically
equilibrated yet. This can be seen by plotting the evolution of the total particle number density
$n_\tinys{tot}(t)=\int_\mathbf{k} n_\mathbf{k}(t)$, as shown in Fig.~\ref{figure3}. 
One observes that the kinetically pre-equilibrated state remembers the initial particle number, and chemical equilibration occurs at a very
slow rate.
\par
\begin{figure}
	\centerline{\includegraphics[clip=true,width=.45\textwidth]{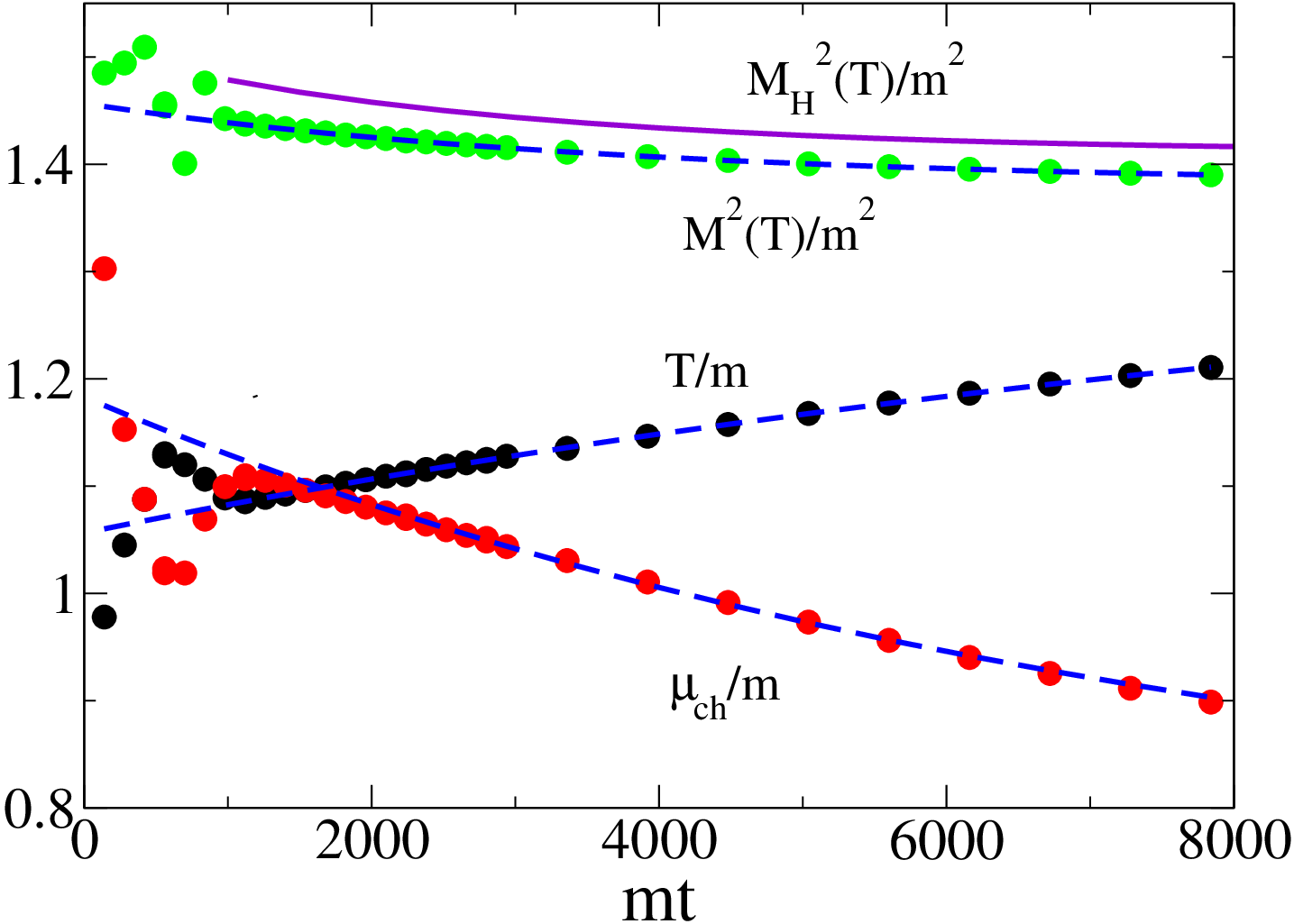}}
\caption{Evolution of effective mass, temperature and chemical potential (T1 case). Hartre mass $M_H$ is included for
comparison.}
\label{figure4}
\end{figure}

Fitting the distribution $n_{\mathbf{k}}(t)$ and frequency $\omega_{\mathbf{p}}(t)$ with an equilibrium
form we can extract an effective temperature $T_\tinys{eff}(t)$, chemical potential $\mu_\tinys{eff}(t)$ and mass
$M_\tinys{eff}(t)$ (see Fig.~\ref{figure4}). Exponential fits for the effective temperature and chemical potential indicate that
final equilibrium is reached at $mt\sim 10^{4-5}$. Comparing to the case in 2+1 dimensions \cite{Juchem:2003bi}, chemical
equilibration seems to be much slower. The values of the effective masses seem to be roughly the same in the Basketball and
Hartree approximations (also included in Fig.~\ref{figure4}).

\section{Broken Phase}
In the broken phase $\phi \neq 0$ and hence one can compare the approach to equilibrium between the two-loop and
Basketball approximations. 
This is particularly interesting for the two-loop case, since perturbation theory predicts no on-shell scattering for
the corresponding self-energy contribution. Thus a study of the time evolution using, for instance, Boltzmann-like equations will fail to
show equilibration. An approach based on the 2PI effective action, however, includes higher-order self-energy
contributions that do contain on-shell scattering. It also takes into account off-shell scattering, which can play an
important role far away from equilibrium. 
\par
\begin{figure}
	\begin{minipage}{.5\textwidth}
		\includegraphics[clip=true,width=.9\textwidth]{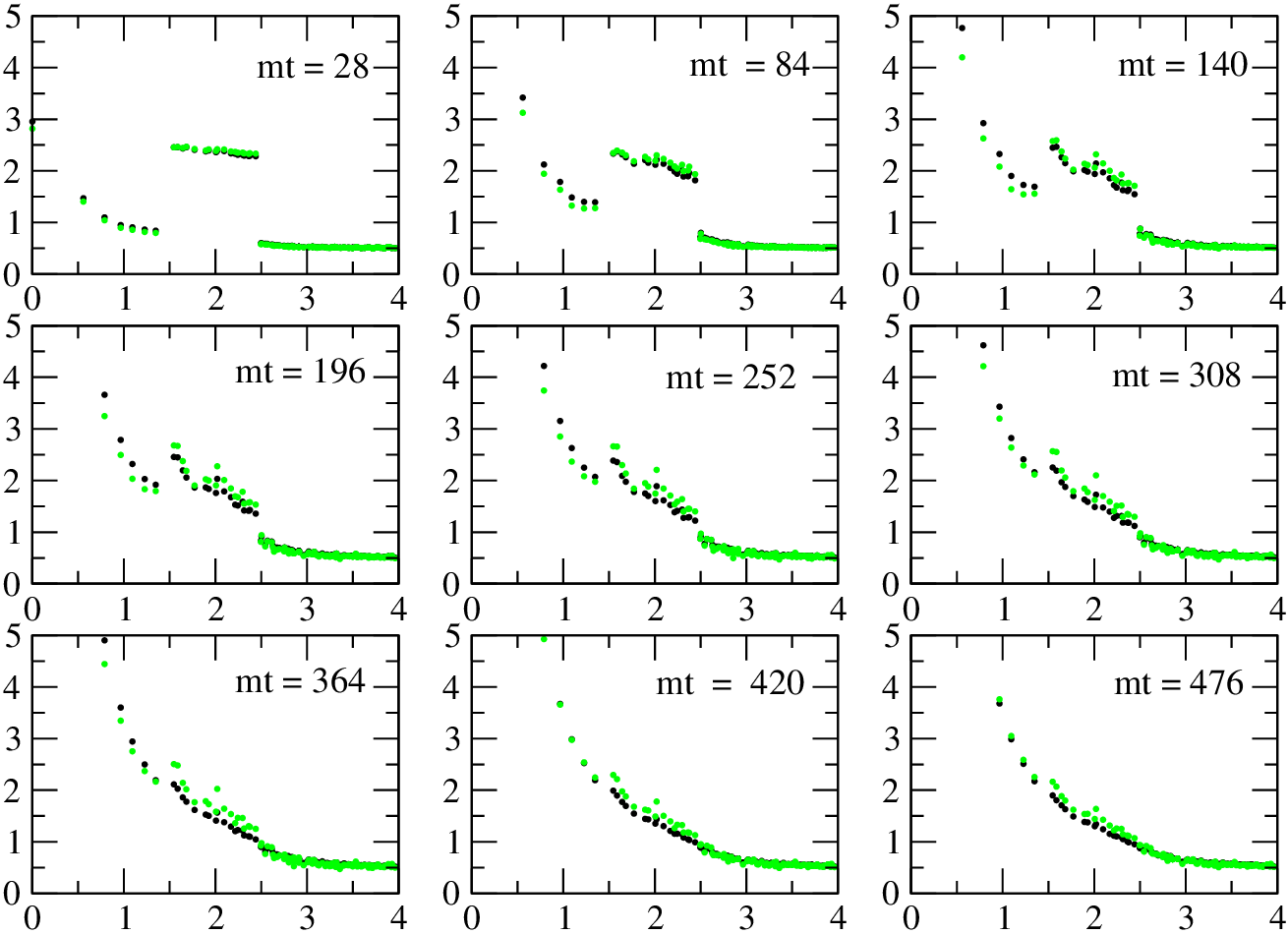}
	\end{minipage}
	\begin{minipage}{.5\textwidth}
		\includegraphics[clip=true,width=.9\textwidth]{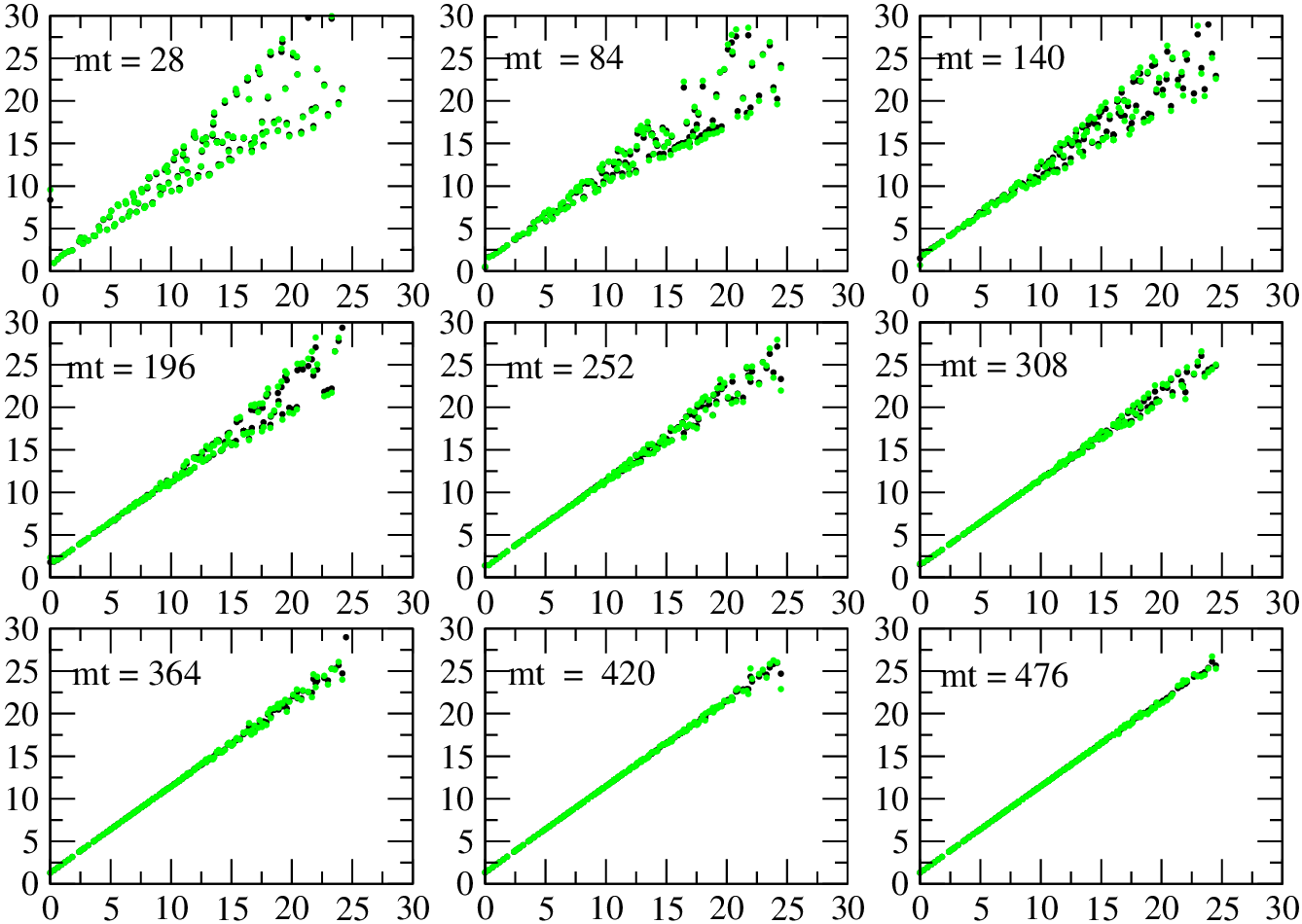}
	\end{minipage}
\caption{Evolution in time of the occupation numbers, $n_{\mathbf{k}}$
\textit{vs.~}$\omega_{\mathbf{k}}$ (left), and dispersion relation $\omega_{\mathbf{k}}^2$
vs.~${\mathbf{k}}^2$ (right), for the T1 initial condition. We compare the results of the
two-loop
(green/gray) and the Basketball approximation
(black).}
\label{figure5}
\end{figure}
\begin{figure}
	\centerline{\includegraphics[clip=true,width=.5\textwidth]{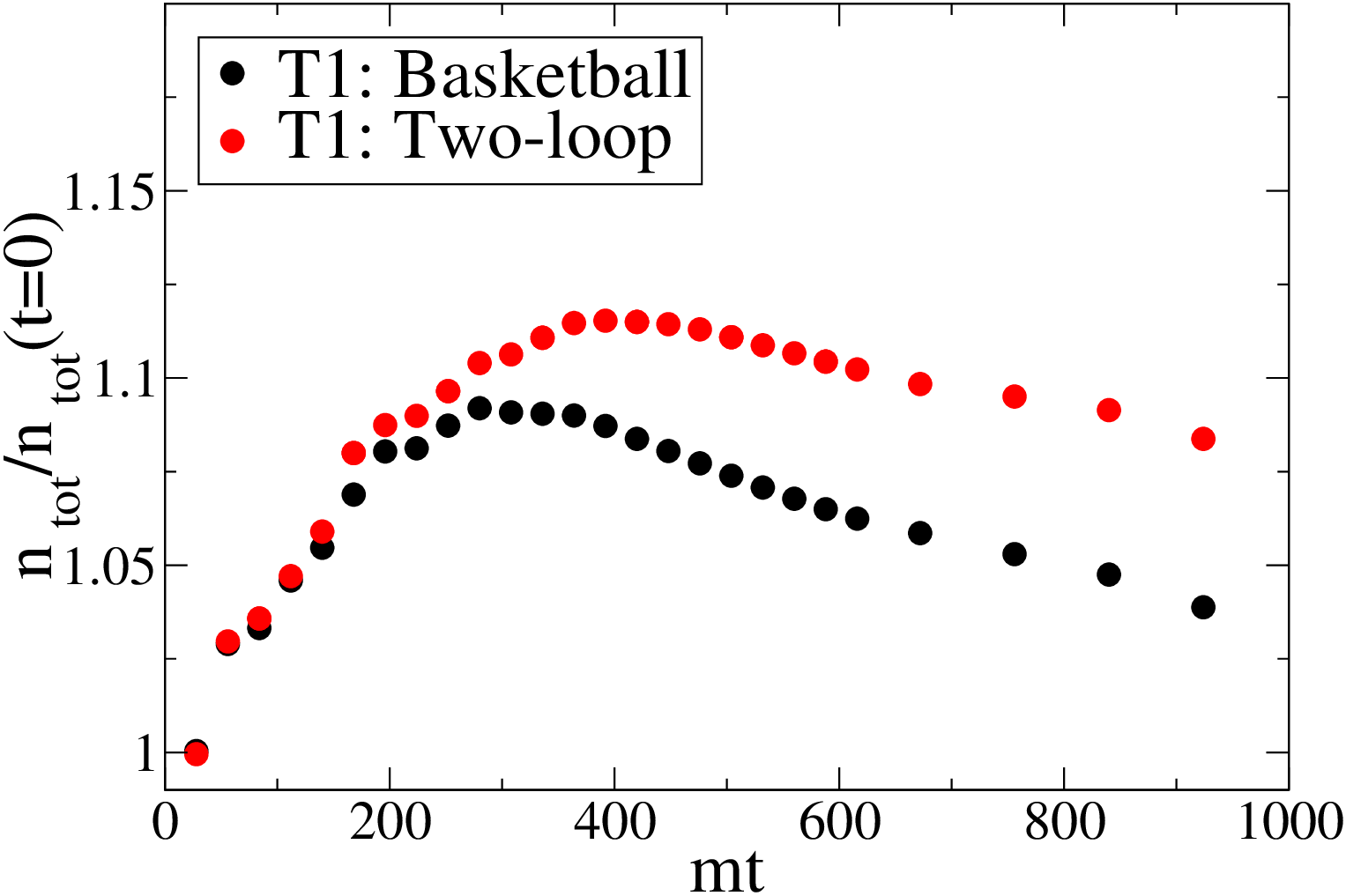}}
\caption{Evolution of total particle number density $n_\tinys{tot}$ for the two-loop and Basketball approximations.}
\label{figure6}
\end{figure}

The system is initialized with a distribution function of the form \eref{distro}. The mean field is taken at the tree
level vacuum expectation value $\phi(t=0)=v_\tinys{tree}=\sqrt{6|m^2|/\lambda}$. For not so large couplings this is
close to the actual variational
value, so the time evolution of $\phi(t)$ should not affect much the dynamics of the two-point functions that we
use to monitor the process of equilibration. In our simulations we use $\lambda=1$.\footnote{This requires a longer
memory kernel. We cut the memory at $mt_\tinys{cut}=84$.} The evolution of the distribution
function and dispersion relation defined 
according to \eref{particlenumber} and \eref{particleenergy} is shown in Fig.~\ref{figure5}.
Surprisingly, the process of equilibration in the two-loop approximation seems to occur almost as fast as in the
Basketball case. The same holds for the chemical equilibration that takes place at later stages of the evolution,
as one can see in Fig.~\ref{figure6}. As in the symmetric case, chemical equilibration is very slow, which impedes the
approach to complete equilibrium.
\section{Conclusions}
In this work several truncations of the 2PI effective action have been used to study equilibration in $\varphi^4$
theory in 3+1 dimensions. Early in the evolution, the distribution function and dispersion relation stabilize and take
equilibrium-like forms due to a relatively fast kinetic equilibration (see also \cite{Berges:2004ce}). This occurs for both the two-loop and Basketball
approximations, which take scattering into account. Surprisingly, equilibration is equally rapid in both approximations.
An analysis of damping rates of distrubances close to equilibrium leads to a similar conclusion \cite{Arrizabalaga:2005tf}. 
\par
The pre-equilibrated state has not completely lost the memory of the initial state. In particular, initial conditions
with different total particle number pre-equilibrate to different states. The later approach to complete equilibrium is
very slow and mainly determined by the time scales associated with chemical equilibration.
\section*{Acknowledgements}
I would like to thank J. Smit and A. Tranberg for collaboration on this work and the organizers for an enjoyable meeting. This work is supported by FOM.

\end{document}